\documentclass[a4paper,11pt]{article}
\usepackage[hang,flushmargin]{footmisc}
\usepackage{amsmath}
\usepackage{amsthm}
\usepackage{amssymb}
\usepackage{verbatim}
\usepackage{enumitem}
\usepackage{multirow}
\usepackage[protrusion=true,expansion=true]{microtype}
\usepackage{lmodern}
\usepackage{fullpage} 
\usepackage[authoryear]{natbib}
\usepackage{longtable}
\usepackage{booktabs} 
\usepackage{csquotes}
\usepackage{tikz}
\usetikzlibrary{arrows.meta,fit,positioning,shapes}
\usepackage[margin=1cm, bf]{caption}
\usepackage{parskip}
\usepackage{hyperref}

\newcommand{\idx}[1]{\texttt{[}\/#1\/\texttt{]}}
\DeclareMathOperator{\concat}{\operatorname{\oplus}}

\setlist[itemize]{label={---}}
\renewcommand*{\descriptionlabel}[1]{\hspace\labelsep
\parbox{\linewidth}{#1}}
\newlist{labelledlist}{description}{2}
\setlist[labelledlist]{style=nextline, leftmargin=2em, before=\renewcommand*{\descriptionlabel}[1]{\hspace\labelsep}}
\bibpunct{(}{)}{,}{a}{}{,}
\hypersetup{colorlinks=true, allcolors=blue}
\title{Non-local redundancy:\\Erasure coding and dispersed replicas for robust retrieval\\in the Swarm peer-to-peer network}

\author{Viktor Trón, Viktor Tóth, Callum Toner, Dan Nickless,\\ Dániel A.~Nagy, Áron Fischer \& György Barabás$^\ast$\\
\\
$^\ast$\small{Division of Biology, Dept.~IFM, Link\"oping University, SE-58183 Link\"oping, Sweden}
}
\date{}

\usepackage[titles]{tocloft}
\begin{document}

\maketitle

\vspace{1cm}

\begin{abstract}
\noindent This paper describes in detail how erasure codes are implemented in the Swarm system. First, in Section \ref{sec:error-correcting-codes}, we introduce erasure codes, and show how to apply them to files in Swarm (Section \ref{sec:erasure}). In Section  
\ref{sec:levels}, we introduce security levels of data availability and derive their respective parameterisations. In Section \ref{sec:dispersed-replicas}, we describe a construct that enables cross-neighbourhood redundancy for singleton chunks and which completes erasure coding. Finally, in \ref{sec:strategies}, we propose a number of retrieval strategies applicable to erasure-coded files.

\vspace{1cm}

{\small
\tableofcontents
}
\end{abstract}
\newpage

\section{Error correcting codes}\label{sec:error-correcting-codes}
%

\emph{Error correcting codes} are widely utilised in the context of data storage and transfer to ensure data integrity even in the face of a system fault. Error correction schemes define how to rearrange the original data by adding redundancy to its representation before upload or transmission (\emph{encoding}\,) so that it can correct corrupted data or recover missing content upon retrieval or reception (\emph{decoding}\,). The different schemes are evaluated by quantifying their strength (\emph{tolerance}, in terms of the rate of data corruption and loss) as a function of their cost (\emph{overhead}, in terms of storage and computation).

In the context of  computer hardware architecture, synchronising arrays of disks is crucial for providing resilient storage in data centres.
In \emph{erasure coding}, %
%
%
in particular, the problem can be framed as follows: How does one encode the stored data into shards distributed across the disks so that the data remains fully recoverable in the face of an arbitrary probability that any one disk becomes faulty?
Similarly, in the context of Swarm's distributed immutable chunk store, the problem can be reformulated as follows: How does one encode the stored data into chunks distributed across neighbourhoods in the network so that the data remains fully recoverable in the face of an arbitrary probability that any one chunk is not retrievable?%
\footnote{%
We will assume that the retrieval of any one chunk fails with equal and independent probability.}

Reed-Solomon coding (RS)  \citep{reed1960polynomial,lubyetal1995CRS,plank2006optimizing,li2013erasure}
is the father of all  error correcting codes and also the most widely used and implemented.%
\footnote{%
For a thorough comparison of an earlier generation of implementations of RS, see \citet{plank2009performance}.}
When applied to data of $m$ fixed-size blocks (message of length $m$), it produces an encoding of $m+k$ \emph{codewords} (blocks of the same size)
in such a way that having any $m$ out of $m+k$ blocks is enough to reconstruct the original data. Conversely, $k$ puts an upper bound on the number of \emph{erasures} allowed (number of blocks unavailable) for full recoverability, i.e., it expresses (the maximum) \emph{loss tolerance}.%
\footnote{Error correcting codes that have a focus on correcting data loss are referred to as \emph{erasure codes}, a typical scheme of choice for distributed storage systems \citep{balaji2018erasure}.}
$k$ is also the count of \emph{parities}, quantifying the data blocks added during the encoding on top of the original volume, i.e., it expresses \emph{storage overhead}. While RS is, therefore, optimal for storage (since loss tolerance cannot exceed the storage overhead),
it has high bandwidth demands%
\footnote{Both the encoding and the decoding of RS codes takes $O(mk)$ time (with $m$ data chunks and $k$ parities). However, we found computational overhead both insignificant for a network setting as well as undifferentiating.}
for local repair processes.%
\footnote{Entanglement codes \citep{estrada2018alpha, estrada2019building} require a minimal bandwidth overhead for a local repair, but at the cost of storage overhead that is in multiples of $100\%$.}
The decoder needs to retrieve $m$ chunks to recover a particular unavailable chunk.
Hence, ideally, RS is used on files which are supposed to be downloaded in full,%
\footnote{Or in  fragments large enough  to include the data span over which the encoding is defined, such as videos.}
 but it is inappropriate for use cases needing only local repairs.%
\footnote{Use cases requiring random access to small amounts of data (e.g., path lookup) benefit from simple replication to optimise on bandwidth, which is suboptimal in terms of storage \citep{weatherspoon2002erasure}.}

When using RS, it is customary to use \emph{systematic} encoding, which means that the original data forms part of the encoding, i.e., the parities are actually added to it.%
\footnote{Our library of choice implementing exactly such a scheme is \url{https://github.com/klauspost/reedsolomon}.}
\newpage
\section{Erasure coding in the Swarm hash tree}
\label{sec:erasure}

\begin{figure}[!ht]
   \centering
   \begin{tikzpicture}
\node[draw,dashed] (root) at (5,3) {root hash};
\node[draw,dashed] (h1) at (1,1) {$h_1$};
\node[draw,dashed] (h2) at (2,1) {$h_2$};
\node[draw,dashed] (h3) at (3,1) {$h_3$};
\node (dots) at (5,1) {$\cdots$};
\node[draw,dashed] (h128) at (7,1) {$h_{128}$};
\node[draw,fit=(h1) (h2) (h3) (dots) (h128)]{};
\draw (root) -- (h1);
\draw (root) -- (h2);
\draw (root) -- (h3);
\draw (root) -- (h128);
\node[draw,dashed] (g1) at (-1.4,-1) {$h^1_1$};
\node[draw,dashed] (g2) at (-0.6,-1) {$h^1_2$};
\node (gdots) at (0,-1) {$\cdots$};
\node[draw,dashed] (g128) at (0.7,-1) {$h^1_{128}$};
\node[draw,fit=(g1)(g2)(gdots)(g128)]{};
\draw (h1) -- (g1);
\draw (h1) -- (g2);
\draw (h1) -- (g128);
\node[draw,dashed] (f1) at (1.8,-1) {$h^2_1$};
\node[draw,dashed] (f2) at (2.6,-1) {$h^2_2$};
\node (fdots) at (3.2,-1) {$\cdots$};
\node[draw,dashed] (f128) at (3.9,-1) {$h^2_{128}$};
\node[draw,fit=(f1)(f2)(fdots)(f128)]{};
\draw (h2) -- (f1);
\draw (h2) -- (f2);
\draw (h2) -- (f128);
\node (moredots) at (5,-1) {$\cdots$};
\node[draw] (c1) at (-2.3,-3) {chunk 1};
\node[draw] (c2) at (0,-3) {chunk 2};
\node at (1.4,-3) {$\cdots$};
\node[draw] (c129) at (3,-3) {chunk 129};
\node (cdots) at (4.5,-3) {$\cdots$};
\node[draw] (cn) at (6,-3) {chunk N};
\draw (g1) -- (c1);
\draw (g2) -- (c2);
\draw (f1) -- (c129);
\end{tikzpicture}
   \caption[Swarm hash split]{The Swarm tree is the data structure encoding how a document is split into chunks.}
   \label{fig:Swarm-hash-split}
\end{figure}
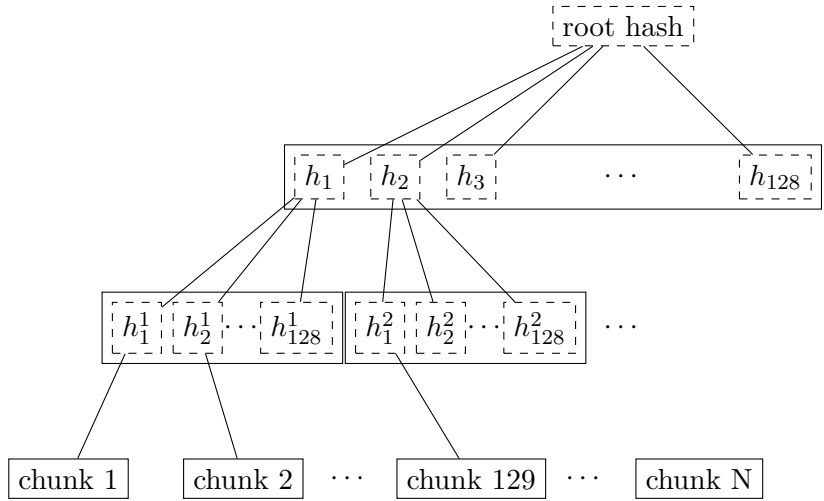

Swarm uses the \emph{Swarm hash tree} to represent files. This structure is a Merkle tree \citep{merkle1980protocols}, whose leaves are the consecutive segments of the input data stream. These segments are turned into chunks and are distributed among the Swarm nodes for storage. The consecutive chunk references (either in the form of an address or an address and an encryption key) are written into a chunk at a higher level.
These so-called \emph{packed address chunks} (PACs) constitute the intermediate chunks of the tree.
The branching factor $b$ is chosen so that the references to its children fill up a full chunk.
With a reference size of 32 or 64 (hash size 32) and a chunk size of 4096 bytes, the value of $b$ is 128 for unencrypted, and 64 for encrypted content
(Figure \ref{fig:Swarm-hash-split}).

Note that on the right edge of the hash tree, the last chunk of each level may be shorter than 4K: in fact, unless the file is exactly $4\cdot b^n$ kilobytes long, there is always at least one \emph{incomplete chunk}.  Importantly, it makes no sense to wrap a single chunk reference in a PAC, so it is attached to the first level where there are open chunks. Such \emph{"dangling" chunks} will appear if and only if the file has a zero digit in its $b$-ary representation.

During file retrieval, a Swarm client starts from the root hash reference and retrieves the corresponding chunk. Interpreting the metadata as encoding the span of data subsumed under the chunk, it decides that the chunk is a PAC if the span exceeds the maximum chunk size.
In case of standard file download, all the references packed within the PAC are followed, i.e., the referenced chunk data is retrieved.

\begin{figure}[hp]
   \centering
   \resizebox{1\textwidth}{!}{
        \begin{tikzpicture}[
level/.style={sibling distance=15mm, line width=0.6pt, level distance=18mm},
hash/.style={fill=white, rounded corners=2pt,draw,minimum size=1.2cm},
phash/.style={fill=lightgray!50, rounded corners=2pt,dotted,draw,minimum size=1.2cm},
chunk/.style={fill=lightgray, rounded corners=2pt,draw,minimum size=1.2cm},
midchunk/.style={draw=none,minimum size=0.2cm}
]

\node [hash] (root) {$H$}
  child[grow=down,draw=none] { node {} edge from parent[<-,shorten >=12pt]}
  child[grow=right,draw=none,level distance=5cm] { node (swh) {Swarm root hash} edge from parent[draw=none] }
  child {node [hash] (n-10) {$H_{0}$} edge from parent[draw=none]
    child[grow=down,draw=none] { node {} edge from parent[<-, shorten >=12pt]}
    child {node [hash] (n-2l0) {$H_{0}$} edge from parent[draw=none]
      child[thick] {node [midchunk] (2l0) {} edge from parent[draw=none]
        child {node [hash] at (0,1.2) (1l0) {$H_{0}$} edge from parent[draw=none]
          child[grow=down,draw=none] { node {} edge from parent[<-, shorten >=12pt]}
          child {node [hash] (0l0) {$H_{0}$} edge from parent[<-, draw=none]
            child {node [chunk] {$C_{0}$}}
          }
          child {node [hash] (0l1) {$H_{1}$} edge from parent[<-, draw=none]
            child {node [chunk] (cl1) {$C_{1}$}}
          }
          child[missing]
          child {node [hash] (0ll) {$H_{111}$} edge from parent[<-, draw=none]
            child {node [chunk] (cll) {}}
          }
          child { node [phash] (p1-0ll) {$P_{0}$} edge from parent[draw=none]}
          child[missing]
          child { node [phash] (p15-0ll) {$P_{15}$} edge from parent[draw=none]}
        }
        child {node [hash] at (0,1.2) (1l1) {$H_{1}$} edge from parent[draw=none]
          child[thick,loosely dotted, shorten >=6mm, thick,<-] {node {}}
        }
        child[missing]
        child {node [hash] at (0,1.2) (1ll) {$H_{111}$} edge from parent[draw=none]
          child[thick,loosely dotted, shorten >=6mm, thick,<-] {node {}}
        }
        child { node [phash] at (0.2,1.2) (p1l) {$P_{0}$} edge from parent[draw=none]}
        child[missing]
        child { node [phash] at (0,1.2) (p1l0) {$P_{15}$} edge from parent[draw=none]}
      }
    }
    child {node [hash] (n-2l1) {$H_{1}$} edge from parent[draw=none]
      child[thick,loosely dotted, shorten >=6mm, thick,<-] {node {}}
    }
    child[missing]
    child {node [hash] (n-2ll) {$H_{111}$} edge from parent[draw=none]
      child[thick,loosely dotted, shorten >=6mm, thick,<-] {node {}}
    }
    child { node [phash] at (0.2,0) (p1-2ll) {$P_{0}$} edge from parent[draw=none]}
    child[missing]
    child { node [phash] (p15n-2ll) {$P_{15}$} edge from parent[draw=none]}
    child[missing]
    child[missing]
  }
  child {node [hash] (n-11) {$H_{1}$} edge from parent[draw=none]
    child[thick,loosely dotted, shorten >=6mm, thick,<-] {node {}}
  }
  child[missing]
  child[missing]
  child[missing]
  child[missing]
  child { node [hash] (n-1l) {$H_{111}$} edge from parent[draw=none]
    child[grow=down,draw=none] { node {} edge from parent[<-, shorten >=12pt]}
    child[missing]
    child {node [hash] (n-2r0) {$H_{0}$} edge from parent[draw=none]
      child[thick,loosely dotted, shorten >=6mm, thick,<-] {node {}}
    }
    child {node [hash] (n-2r1) {$H_{1}$} edge from parent[draw=none]
      child[thick,loosely dotted, shorten >=6mm, thick,<-] {node {}}
    }
    child[missing]
    child {node [hash] (n-2rl) {$H_{111}$} edge from parent[draw=none]
      child[grow=down] {node [midchunk] (2rl) {} edge from parent[draw=none]
        child {node [hash] at (0,1.2) (1r0) {$H_{0}$} edge from parent[draw=none]
          child[thick,loosely dotted, shorten >=6mm, thick,<-] {node {}}
        }
        child {node [hash] at (0,1.2) (1r1) {$H_{1}$} edge from parent[draw=none]
          child[thick,loosely dotted, shorten >=6mm, thick,<-] {node {}}
        }
        child[missing]
        child {node [hash] at (0,1.2) (1rl) {$H_{111}$} edge from parent[draw=none]
          child[grow=down,draw=none] { node {} edge from parent[<-, shorten >=12pt]}
          child {node [hash] (0r0) {$H_{0}$} edge from parent[<-, draw=none]
            child {node [chunk] (cr0) {}}
          }
          child {node [hash] (0r1) {$H_{1}$} edge from parent[<-, draw=none]
            child {node [chunk] (cr1) {}}
          }
          child[missing]
          child {node [hash] (0rl) {$H_{111}$} edge from parent[<-, draw=none]
            child {node [chunk] (crl) {$C_{m}$}
            }
          }
          child { node [phash] at (0.2,0) (p100) {$P_{0}$} edge from parent[draw=none]}
          child[missing]
          child { node [phash] (p150) {$P_{15}$} edge from parent[draw=none]}
        }
        child { node [phash] at (0.2,1.2) (p11) {$P_{0}$} edge from parent[draw=none]}
        child[missing]
        child { node [phash] at (0,1.2) (p151) {$P_{15}$} edge from parent[draw=none]}
      }
    }
    child { node [phash] at (0.2,0) (p1n-2) {$P_{0}$} edge from parent[draw=none]}
    child[missing]
    child { node [phash] (p15n-2) {$P_{15}$} edge from parent[draw=none]}
  }
  child { node [phash] at (0.2,0) (p1n) {$P_{0}$} edge from parent[draw=none]}
  child[missing]
  child { node [phash] (p15n-11) {$P_{15}$} edge from parent[draw=none]};

\path (p1n) -- (p15n-11) node [midway,font=\large] {$\ldots$};
\path (p1n-2) -- (p15n-2) node [midway,font=\large] {$\ldots$};
\path (p11) -- (p151) node [midway,font=\large] {$\ldots$};
\path (p100) -- (p150) node [midway,font=\large] {$\ldots$};
\path (p1l) -- (p1l0) node [midway,font=\large] {$\ldots$};
\path (p1-0ll) -- (p15-0ll) node [midway,font=\large] {$\ldots$};
\path (p1-2ll) -- (p15n-2ll) node [midway,font=\large] {$\ldots$};

\path (n-11) -- (n-1l) node [midway,font=\large] {$\ldots$};
\path (n-2l0) -- (2l0) node [midway,font=\large,sloped] {$\ldots$};
\path (n-2rl) -- (2rl) node [midway,font=\large,sloped] {$\ldots$};
\path (1l1) -- (1ll) node [midway,font=\large] {$\ldots$};
\path (1r1) -- (1rl) node [midway,font=\large] {$\ldots$};
\path (p1l0) -- (1r0) node [midway,font=\large] {$\ldots$};
\path (n-2l1) -- (n-2ll) node [midway,font=\large] {$\ldots$};
\path (n-2r1) -- (n-2rl) node [midway,font=\large] {$\ldots$};
\path (0l1) -- (0ll) node [midway,font=\large] {$\ldots$};
\path (0r1) -- (0rl) node [midway,font=\large] {$\ldots$};

\begin{scope}[shorten >=.5cm,thin]
\draw [->] (swh) -> (root);
\end{scope}

\node[rounded corners=2pt, draw=black, minimum height=1.1cm, fit=(n-10) (p15n-11)] {};

\node[rounded corners=2pt, draw=black, minimum height=1.1cm, fit=(0l0) (p15-0ll)] {};
\node[rounded corners=2pt, draw=black, minimum height=1.1cm, fit=(n-2l0) (p15n-2ll)] {};
\node[rounded corners=2pt, draw=black, minimum height=1.1cm, fit=(n-2r0) (p15n-2)] {};
\node[rounded corners=2pt, draw=black, minimum height=1.1cm, fit=(1l0) (p1l0)] {};
\node[rounded corners=2pt, draw=black, minimum height=1.1cm, fit=(1r0) (p151)] {};

\node[rounded corners=2pt, draw=black, minimum height=1.1cm, fit=(0r0) (p150)] {};

\end{tikzpicture}
   }
   \caption[Swarm hash erasure]{The Swarm tree with extra parity chunks using $m=112$ out of $n=128$ RS encoding. Chunks $P_{0}$ through $P_{15}$ are parity data for chunks $H_0 $ through $H_{111}$ on every level of intermediate chunks.}
   \label{fig:Swarm-hash-erasure}
\end{figure}
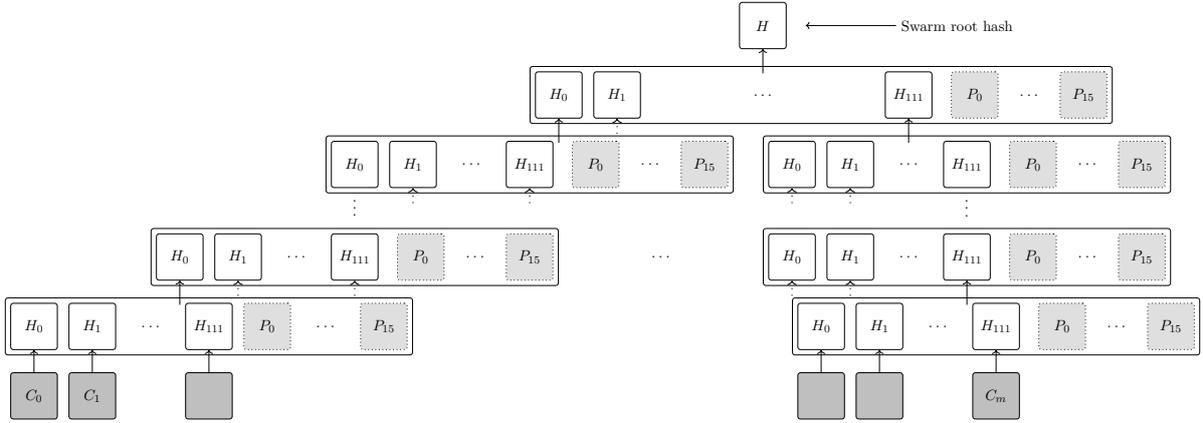

PACs offer a natural and elegant way to achieve consistent redundancy within the Swarm hash tree.
The input data for an instance of erasure coding is the chunk data of the children, with the equal-sized bins corresponding to the chunk data of the consecutive references packed into it. The idea is that instead of having each of the $b$ references packed represent children, only $m$ would, and the rest of the $k=b-m$ would encode RS parities (see Figure \ref{fig:Swarm-hash-erasure}).

The \emph{chunker} algorithm that incorporates PAC-scoped RS encoding would work as follows:
\begin{enumerate}[noitemsep]
\item Set the input to the actual data level and produce a sequence of chunks from the consecutive 4K segments of the data stream. Choose $m$ and $k$ such that $m+k=b$ is the branching factor (128 for unencrypted, and 64 for encrypted content).
\item Read the input one chunk at a time. Count the chunks by incrementing a counter $i$.
\item Repeat Step 2 until either $i = m$ or there is no more data left.
\item Use the RS scheme on the last $i\leq m$ chunks to produce $k$ parity chunks resulting in a total of $n = i+k \leq b$ chunks.
\item Concatenate the references of all these chunks to result in a packed address chunk (of size $h\cdot n$) on the level above. If this is the first chunk on that level, set the input to this level and spawn this same procedure from Step 2.
\item When the input is consumed, signal the end of input to the next level and quit the routine. If there is no next level, record the single chunk as the root chunk and use the reference to refer to the entire file.
\end{enumerate}

This pattern repeats itself all the way down the tree. Thus, hashes $H_{m+1}$ through $H_{127}$ point to parity data for chunks pointed to by $H_0$ through $H_{m}$.%
\footnote{Since parity chunks $P_i$ do not have children, the tree structure does not have uniform depth.}

\section{Levels of security and the number of parities}\label{sec:levels}

Non-local redundancy is presented here as a scheme of encoding that allows strategies of retrieval in order to guarantee data availability.
With packed address chunks set as the scope of erasure codes, it is crucial that we use the right number of shards and parities among the children of an intermediate node in the Swarm hash tree representing a file.      
Given assumptions about chunk retrieval error rates and the number of parities used, one can calculate the degree of certainty that the data can be recovered without error. One can even apply the same logic backwards: given some level of certainty with which we want recovery to be error-free, we can compute how many parities should be used to provide that level of safety.
In what follows, we give a formal exposition of how to find these parity counts.

Let there be $m$ original chunks and $k$ parity chunks, such that any $m$ chunks out of the total $n = m + k$ ones are fully recoverable after the loss of any $k$ of them. In the process of retrieving the $n$ chunks, what is the likelihood of overall data corruption, given a per-chunk probability of error $\epsilon$?

By ``overall data corruption'', we mean that more than $k$ chunks are damaged in the data retrieval process. We assume that each chunk's probability of error is independent of other chunks. In that case, the problem boils down to the independent drawing of $n$ chunks, each of which undergo a \emph{Bernoulli trial} of being faulty with probability $\epsilon$. The total number of faulty chunks out of $n$ independent Bernoulli trials is given by the \emph{binomial distribution}:
\begin{equation}
  B(i, n, \epsilon) = \binom{n}{i} \epsilon^k (1-\epsilon)^{n-i} .
  \label{eq-PMF-binomial}
\end{equation}
This expression is the probability mass function for the binomial distribution, yielding the probability that out of $n$ chunks, exactly $i$ will be faulty---assuming that the per-chunk probability of error is $\epsilon$.

Since there are $k$ parities out of the $n$ chunks, the system can tolerate up to $k$ chunk errors. The probability that no more than $k$ errors accumulate can be expressed by summing Equation~\ref{eq-PMF-binomial} over $i$ up to $k$:
\begin{equation}
  P(k, n, \epsilon) = \sum_{i=0}^k \binom{n}{i} \epsilon^k (1-\epsilon)^{n-i} ,
  \label{eq-CDF-binomial}
\end{equation}
which is the cumulative distribution function of the binomial distribution.

\begin{figure}[!ht]
  \centering
  \includegraphics[width=.7\textwidth]{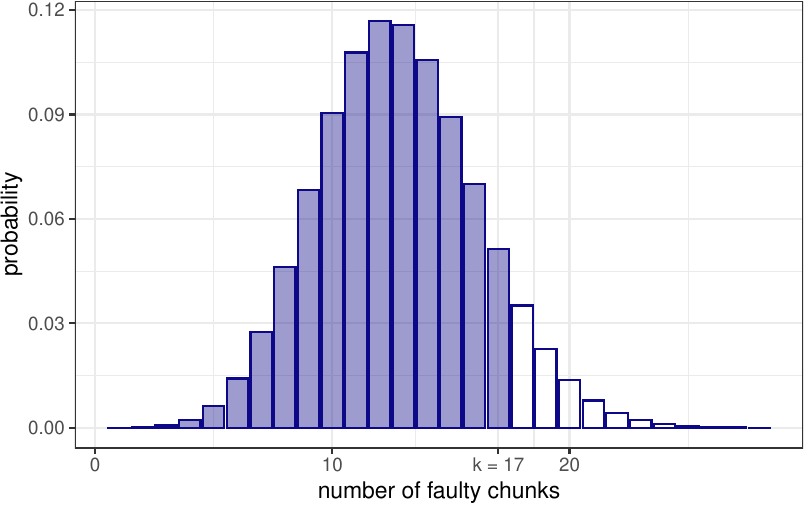}
  \caption{The point at $k=17$ along the binomial distribution, where the probability of exceeding this many errors becomes less than $\alpha = 10\%$. Here, the total number of chunks is $n = 128$, and the per-chunk error rate is $\epsilon = 0.1$.}
  \label{fig:alpha}
\end{figure}

One typical question is the following: given the number of chunks $n$ and a value $\alpha$ such that we want the overall probability of data corruption to be below this value, how many out of the $n$ chunks should be parities? Since $P(k, n, \epsilon)$ is the probability that \emph{no more than} $k$ errors accumulate, $1 - P(k, n, \epsilon)$ is the probability of more than $k$ errors; i.e., that \emph{at least} $k + 1$ errors accumulated and therefore the data are corrupted. We want to keep this probability below $\alpha$, so we can write
\begin{equation}
  \alpha \ge 1 - P(k, n, \epsilon) .
  \label{eq-CDF-cond}
\end{equation}
Rearranging, we have
\begin{equation}
  1 - \alpha \le P(k, n, \epsilon) .
  \label{eq-CDF-cond2}
\end{equation}
That is, we are looking for values of $k$ which will satisfy this inequality (Figure \ref{fig:alpha}). This can be obtained by inverting the cumulative distribution function in $k$, resulting in the quantile function $Q(1 - \alpha, n, \epsilon)$. While this inverse has no convenient closed-form expression, it can be efficiently evaluated numerically for any set of input parameters. As with any cumulative distribution function, $P(k, n, \epsilon)$ is monotonically increasing in $k$. Applying the inverse on both sides of Equation~\ref{eq-CDF-cond2} therefore does not flip the direction of the inequality, and gives $k \ge Q(1 - \alpha, n, \epsilon)$. Or if we look for the smallest $k$ satisfying this condition:
\begin{equation}
  k = Q(1 - \alpha, n, \epsilon) .
  \label{eq-quantile-sol-n}
\end{equation}
This is the formula yielding the minimum number of required parities to keep the overall probability of data corruption below $\alpha$.

\begin{figure}[!ht]
  \centering
  \includegraphics[width=.7\textwidth]{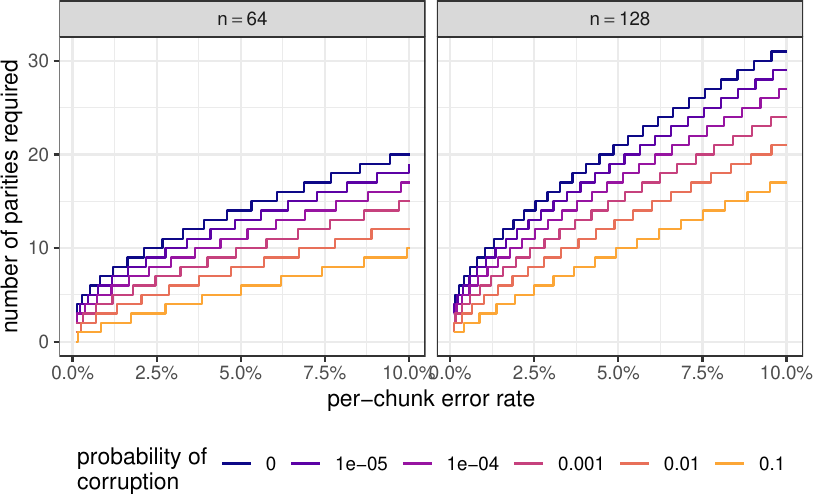}
  \caption{The number of parities needed (ordinate) as a function of the per-chunk error rate $\epsilon$ (abscissa), for keeping the probability of overall data corruption below given limits (colours) and for $n = 64$ chunks (left panel) and $n = 128$ chunks (right panel).}
  \label{fig:perr-lin}
\end{figure}

\begin{figure}[!ht]
  \centering
  \includegraphics[width=.7\textwidth]{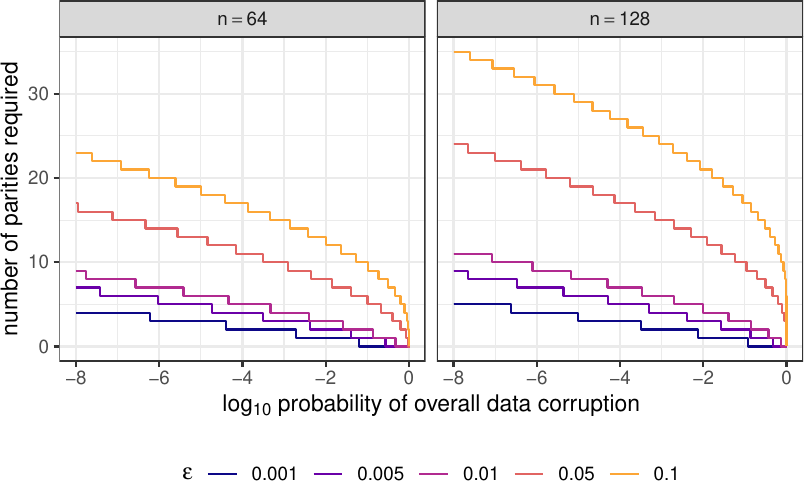}
  \caption{The number of parities required (ordinate) to keep the probability of overall data corruption at a given level (abscissa), for various values of the per-chunk error rate $\epsilon$ (colours) and for $n = 64$ chunks (left panel) and $n = 128$ chunks (right panel).}
  \label{fig:integrity}
\end{figure}

Figure \ref{fig:perr-lin} presents the number of parities needed as a function of error rate for various levels of security.
Figure \ref{fig:integrity}  
presents the number of parities needed to keep the probability of overall data corruption at a given level for various values of the per-chunk error rate.

The same type of problem can also be phrased slightly differently: given a number of chunks $n$, how many parities $k$ should be added to them to keep the overall data corruption probability below some level $\alpha$? In this case, the total number of chunks is $n + k$ (instead of having $n$ chunks, out of which $k$ are parities), and so Equation~\ref{eq-quantile-sol-n} is modified to be
\begin{equation}
  k = Q(1 - \alpha, n + k, \epsilon) .
  \label{eq-quantile-sol}
\end{equation}
While this equation has no closed-form solution for $k$, one can easily find the $k$ satisfying it as long as $k$ is bounded in a relatively small range. In our case, the maximum number of chunks, $n + k$, is 128, and so $k$ is at most $128-n$. This makes it simple to find the value of $k$ compatible with Equation~\ref{eq-quantile-sol}. The number of parities in Tables~\ref{tbl:levels}-\ref{tbl:paranoid} were obtained using this method.

In principle, the exact parity counts can be made user-configurable. However, to make non-local redundancy a transparent and easy-to-use feature, we opted for a simplified yet intuitive interface.
First of all, we set our maximum tolerated error rate of integrity at $10^{-6}$, in other words our security constant expressing our certainty at 6 nines, 99.9999\%.
Second, we propose to use a handful of named security levels of (non-local) redundancy which correspond to assumptions about the maximum error rates of individual chunk retrievals expressed as discrete percentages. 
Table~\ref{tbl:levels} lists the security levels with the  corresponding assumption about the maximum error rate of chunk retrieval. 
\begin{table}[!ht]
  \centering
\begin{tabular}{|c|c|r||r|r|r|r|}
\hline
\multicolumn{2}{|c|}{security}
&\multirow{2}{3.3cm}{\centering error rate\\of chunk retrieval}
&\multicolumn{2}{|c|}{unencrypted}
&\multicolumn{2}{|c|}{encrypted}\\\cline{1-2}\cline{4-7}
level&name&
& chunks & parities 
& chunks & parities 
\\\hline
0     & \textsc{none} &       0\% &   128 &   0 &  64 &   0 \\
1     & \textsc{medium} &     1\% &   119 &   9 &  59 &   9 \\
2     & \textsc{strong} &     5\% &   107 &  21 &  53 &  21 \\
3     & \textsc{insane} &    10\% &    97 &  31 &  48 &  31 \\
4     & \textsc{paranoid} &  50\% &    38 &  90 &  19 &  90 \\
\hline
\end{tabular}
\caption{Security levels for non-local redundancy UI and corresponding assumptions about uniform and independent error rates of individual chunk retrieval. In subsequent columns we specify the composition of full chunks for the security levels for unencrypted (columns 4 and 5) and encrypted (columns 6 and 7) content.}
  \label{tbl:levels}
\end{table}

If the number of file chunks is not a multiple of $m$, it is not possible to proceed with the last batch in the same way as the others. We propose that we encode the remaining chunks with an erasure code that guarantees at least the same level of security as the others.%
\footnote{Note that this is not as simple as choosing the same redundancy. For example, a $50\text{-out-of-}100$ encoding is much more secure against loss than a $1\text{-out-of-}2$ encoding, even though the redundancy is $100\%$ in both cases.}
Overcompensating, we still require the same number of parity chunks even when there are fewer than $m$ data chunks. However, we can also just calculate the necessary parities for all possible incomplete chunks and security levels. Figure~\ref{fig:maintain} plots the number of parities against the number of chunks required:

\begin{figure}[!ht]
  \centering
  \includegraphics[width=.8\textwidth]{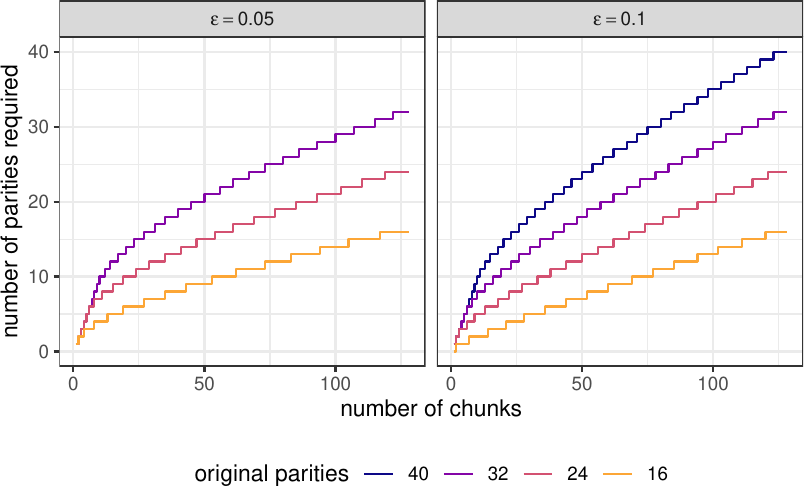}
  \caption{Number of chunks (abscissa) and the corresponding required number of parities (ordinate) such that will maintain the same overall probability of no data corruption as would be the case with 128 chunks, an original number of parities indicated by the colours, and a likelihood $\epsilon$ of an erroneous retrieval of a single chunk indicated in the panel headers.}
  \label{fig:maintain}
\end{figure}

Tables~\ref{tbl:parities} and \ref{tbl:paranoid} show 
the number of chunks that are maintainable for a given number of parities $k$ across various security levels.
Since encrypted chunks are referenced with the hash address followed by the decryption key, an encrypted reference takes up 2 hash-sized segments. Parity chunks added to an encrypted PAC, however, are calculated based on the encrypted shards and are themselves not encrypted, hence their references only use a single hash. 
Thus, the number of effective hash-sized segments 
used is obtained as twice the number of chunks plus the number of parities. Since this can be an odd number and less than 128, in some security levels even the full chunks are not completely full.

\begin{table}[!ht]
\begin{minipage}{.49\linewidth}
\centering
\begin{tabular}{|r|r|r|}
\multicolumn{3}{c}{\textsc{medium}}\\\hline
\multirow{2}{1.5cm}{\centering 
 parities } 
&\multicolumn{2}{|c|}{ chunks }\\\cline{2-3}
&\multicolumn{1}{|c|}{unencrypted}
&\multicolumn{1}{|c|}{encrypted} \\\hline\hline
2 & 1 & - \\
3 & 2-5     & 1-2\\
4 & 6-14    & 3-7\\ 
5 & 15-28   & 7-14\\ 
6 & 29-46   & 14-23\\  
7 & 47-68   & 23-34\\  
8 & 69-94   & 34-47\\  
9 & 95-119  & 47-59\\   
\hline
\multicolumn{3}{c}{\textsc{}}\\
\multicolumn{3}{c}{\textsc{strong}}
\\\hline
4 & 1   &  -\\
5 & 2-3   & 1\\
6 & 4-6   & 2-3\\
7 & 7-10  & 3-5\\
8 & 11-15 & 5-7\\
9 & 16-20 & 8-10\\
10 & 21-26 & 10-13\\
11 & 27-32 & 13-16\\
12 & 33-39 & 16-19\\
13 & 40-46 & 20-23\\
14 & 47-53 & 23-26\\
15 & 54-61 & 27-30\\
16 & 62-69 & 31-34\\
17 & 70-77 & 35-38\\
18 & 78-86 & 39-43\\
19 & 87-95 & 43-47\\
20 &96-104 & 48-52\\
21&105-107 & 52-53\\
\hline
\end{tabular}
\end{minipage}
\begin{minipage}{.49\linewidth}
\centering
\begin{tabular}{|r|r|r|}
\multicolumn{3}{c}{\textsc{}}\\
\multicolumn{3}{c}{\textsc{insane}}\\\hline
\multirow{2}{1.5cm}{\centering 
 parities } 
&\multicolumn{2}{|c|}{ chunks }\\\cline{2-3}
&\multicolumn{1}{|c|}{unencrypted} 
&\multicolumn{1}{|c|}{encrypted} \\\hline\hline
5 & 1     &-   \\
6 & 2     & 1\\
7 & 3     & 1\\ 
8 & 4-5   & 2\\ 
9 & 6-8   & 3-4\\
10 & 9-10  & 4-5\\
11 & 11-13 & 5-6\\
12 & 14-16 & 7-8\\
13 & 17-19 & 8-9\\
14 & 20-22 & 10-11\\
15 & 23-26 & 11-13\\
16 & 27-29 & 13-14\\
17 & 30-33 & 15-16\\
18 & 34-37 & 17-18\\
19 & 38-41 & 19-20\\
20 & 42-45 & 21-22\\
21 & 46-50 & 23-25\\
22 & 51-54 & 25-27\\
23 & 55-59 & 27-29\\
24 & 60-63 & 30-31\\
25 & 64-68 & 32-34\\
26 & 69-73 & 34-36\\
27 & 74-77 & 37-38\\
28 & 78-82 & 39-41\\
29 & 83-87 & 41-43\\
30 & 88-92 & 44-46\\
31 & 93-97 & 46-48\\
\hline
\multicolumn{3}{c}{}\\
\multicolumn{3}{c}{}
\end{tabular}
\end{minipage}
\caption{The number of parities (first column in each table) to be appended to a given number of chunks (second and third column of each table, given as a range) so that the probability of an unsuccessful data retrieval remains below $\alpha = 10^{-6}$. The second column is for unencrypted  chunks, while the third one is for encrypted chunks. The tables are for security levels 1-3, to be continued for security level 4 in Table~\ref{tbl:paranoid}.}
\label{tbl:parities}
\end{table}

\begin{table}[!ht]
\begin{minipage}{.49\linewidth}
\centering
\begin{tabular}{|r|r|r|}
\multicolumn{3}{c}{\textsc{}}\\
\multicolumn{3}{c}{\textsc{paranoid}}\\\hline
\multirow{2}{1.5cm}{\centering 
 parities } 
&\multicolumn{2}{|c|}{ chunks }\\\cline{2-3}
&\multicolumn{1}{|c|}{unencrypted} 
&\multicolumn{1}{|c|}{encrypted} \\\hline\hline
19 & 1 & -\\
23 & 2  & 1\\
26 & 3  & 1\\
29 & 4  & 2\\
31 & 5  & 2\\
34 & 6  & 3\\
36 & 7  & 3\\
38 & 8  & 4\\
40 & 9  & 4\\
43 & 10 & 5\\
45 & 11 & 5\\
47 & 12 & 6\\
48 & 13 & 6\\
50 & 14 & 7\\
52 & 15 & 7\\
54 & 16 & 8\\
56 & 17 & 8\\
58 & 18 & 9\\
59 & 19 & 9\\
\hline
\end{tabular}
\end{minipage}
\begin{minipage}{.49\linewidth}
\centering
\begin{tabular}{|r|r|r|}
\multicolumn{3}{c}{\textsc{}}\\
\multicolumn{3}{c}{\textsc{paranoid} (continued)}\\\hline
\multirow{2}{1.5cm}{\centering 
 parities } 
&\multicolumn{2}{|c|}{ chunks }\\\cline{2-3}
&\multicolumn{1}{|c|}{unencrypted} 
&\multicolumn{1}{|c|}{encrypted} \\\hline\hline
61 & 20 & 10\\
63 & 21 & 10\\
65 & 22 & 11\\
66 & 23 & 11\\
68 & 24 & 12\\
70 & 25 & 12\\
71 & 26 & 13\\
73 & 27 & 13\\
75 & 28 & 14\\
76 & 29 & 14\\
78 & 30 & 15\\
80 & 31 & 15\\
81 & 32 & 16\\
83 & 33 & 16\\
84 & 34 & 17\\
86 & 35 & 17\\
87 & 36 & 18\\
89 & 37 & 18\\
90 & 38 & 19\\
\hline
\end{tabular}
\end{minipage}
\caption{As Table~\ref{tbl:parities}, but for the \textsc{paranoid} security level.}
\label{tbl:paranoid}
\end{table}

As a final note, one should keep in mind that the probability of a failed data retrieval, $\alpha = 10^{-6}$, is not the same as the probability of a failed file retrieval. This is because $\alpha$ is only valid for one 128-chunk segment (64-chunk segment for encrypted content) of a file, not a file as a whole in general. Assuming that retrieval errors may occur independently to any chunk, we can use $\alpha$ and the size of a file to calculate the probability that a file as a whole is successfully retrieved. This probability is $1 - \alpha$ for each 128-chunk segment of a file, so if a file consists of $s$ 128-chunk segments, then the probability is $(1-\alpha)^s$. In terms of bytes: a file of $g$ bytes consists of $g / 2^{12}$ chunks (because $2^{12}$ bytes is 4KB), which then make up for $s = g / (2^{12} \cdot 2^{7})$ 128-chunk segments (because $128 = 2^7$). This means that the probability $P_F$ of a successful file retrieval is
\begin{equation}
  P_F = (1 - \alpha)^{g/2^{19}} ,
  \label{eq-P-file}
\end{equation}
an exponentially decreasing function of the file size $g$. For example, a file of 1GB ($s = 2^{30}$ bytes) with $\alpha = 10^{-6}$ has $P_F = 0.998$, for a failure probability of $1 - P_F = 0.2\%$.

\section{Dispersed replicas}
\label{sec:dispersed-replicas}

This leaves us with only one corner case: it is not possible to use our $m\text{-out-of-}n$ scheme on a single chunk ($m=1$) because it would amount to $k+1$ copies of the same chunk. The problem is that copies of the same chunk all have the same hash and therefore are automatically deduplicated. Whenever a single chunk is left over ($m=1$) (i.e., the root chunk itself), we would need to replicate the chunk in a way that (1) ideally, the replicas are dispersed in the address space in a balanced way, yet (2) their addresses can be known by retrievers who ideally only know the reference to the original chunk's address.

\begin{figure}[!ht]
  \centering
  \includegraphics[width=\textwidth]{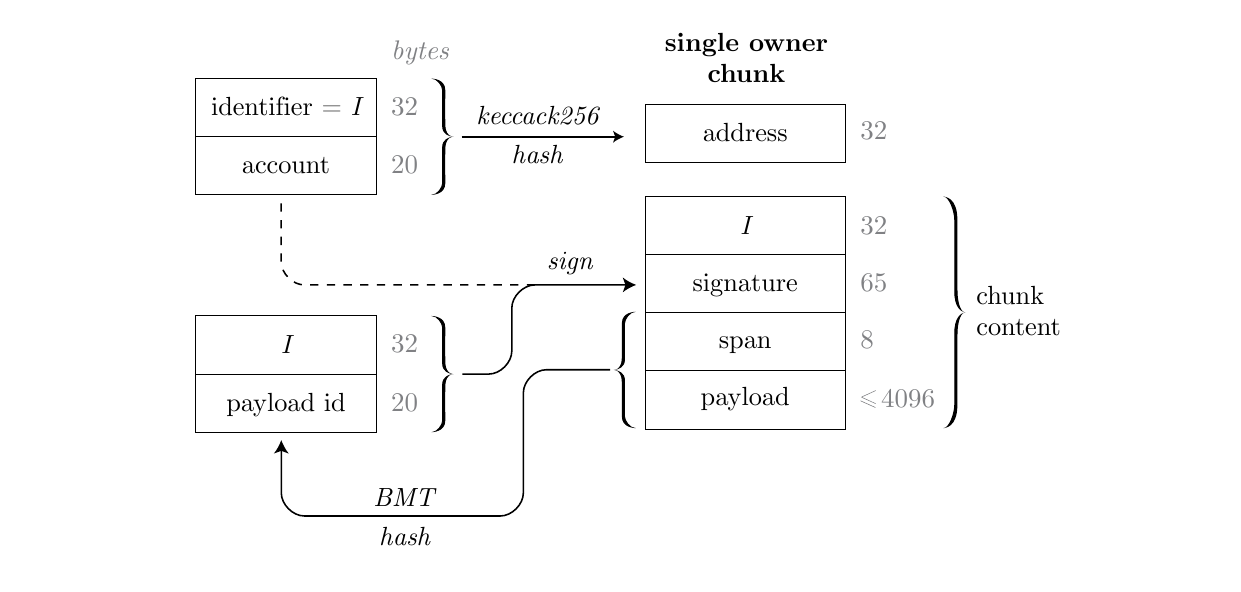}
  \caption{Single owner chunk (SOC). Unlike content-addressed chunks, SOCs obtain their integrity through the signature of their (single) owner and cross-owner immutability through hashing the owner's address in the chunk address (effectively achieving access control via namespacing).}
   \label{fig:soc}
  \end{figure}

Our solution uses Swarm's special construct,  the \emph{single owner chunk} (SOC; Figure~\ref{fig:soc}). Replicas of the root chunk are created by making the chunk data the payload of a number of SOCs. The addresses of these SOCs must be derivable from the original root hash following a deterministic convention shared by uploaders and downloaders.

The address of a SOC is the hash of its ID and the Ethereum address of its owner. In order to create valid SOCs, uploaders need to sign the SOC with the owner's identity, therefore the owner of the SOC must be a consensual identity with their private key publicly revealed.
\footnote{This has the added benefit that third parties can also upload replicas of any chunk.}

The other component of the address, the SOC ID, must satisfy two criteria: (1) it needs to match the payload hash up to 31 bytes and (2) it must provide the entropy needed to mine the overall chunk into a sufficient number of distinct neighbourhoods. (1) is added as a validation criterion for the special case of replica SOCs, while (2) takes care that we can find replicas uniformly dispersed within the address space.
This construct is called \emph{dispersed replica}:

Let us assume $c$ is the content-addressed chunk we need to replicate; $n$ is the number of bits of entropy available to find the nonces that generate  $2^k$  perfectly balanced replicas; initialise a chunk array $\rho$ of length $2^k$ and start with $n$-bit integer $i=0$ and replica counter $C=0$.

\begin{enumerate}[noitemsep]
  \item Create the SOC ID by taking $\mathit{addr}(c)$ and changing the last byte (at index position 31) to  $i$.
  \item Calculate the the SOC address by concatenating ID $id$ and owner $o$%
\footnote{The SOC owner of dispersed replicas has the arbitrary private key $\texttt{0x010...00}$
and the corresponding ether address is 
\texttt{0xdc5b20847f43d67928f49cd4f85d696b5a7617b5}.}
and hash the result using the Keccak256 base hash $a_i:=H(id\concat o)$, and record $c_i=\mathit{SOC}(id,o,c)$.
  \item Calculate the bin this hash belongs to by taking the $k$-bit prefix as big-endian binary number $j$ between $0\leq j<2^k$.
  \item If $\rho\idx{j}$ is unassigned, then let $\rho\idx{j}:=c_i$ and increment $C$.
  \item If $C=2^k$, then quit.
  \item Increment $i$ by one, if $i=2^n$, then quit.
  \item Repeat from Step 1.
\end{enumerate}

With this solution, we are able to provide an arbitrary level of redundancy for the storage of data of any length.
\footnote{Note that if $n$ is small, then generating all $2^k$ balanced replicas may not be achievable, and if $n<k$, this is certainly not possible.
In general, given $n, k$ at least $m$ miss has a probability of $(1 - m/2^k)^{2^n}$.}

Then, depending on the strategy, the downloader can choose which  address to retrieve the chunk from. The obvious choice is the replica closest to the requesting node's overlay address. In other words, the last item of the sorted chunk array $\rho$ using the comparison function:
\begin{equation}
  i<j\Leftrightarrow    	
  \mathit{PO}(    \mathit{Overlay}(node),
  \textsc{Address}(\rho\idx{i}))
  <\mathit{PO}(\mathit{Overlay}(node),
  \textsc{Address}(\rho\idx{j}))
\end{equation}


If the probability of any replica being faulty is $\epsilon$, then, assuming independence, the probability that $n$ parities are faulty is $\epsilon^n$. Here we can write $n = k + 1$; that is, we have one ``original'' chunk and the rest of them are the $k$ parities. Keeping the overall error probability below $\alpha$ then means that
\begin{equation}
  \epsilon^{k+1} = \alpha
  \label{eq-onechunk}
\end{equation}
must be satisfied. Taking logarithms on both sides and rearranging, we get
\begin{equation}
  k = \frac{\log(\alpha)}{\log(\epsilon)} - 1 .
  \label{eq-onechunk-parities}
\end{equation}
This is the number of parities of a singleton chunk required to keep the overall data corruption probability below $\alpha$. The base of the log in Equation~\ref{eq-onechunk-parities} is arbitrary. This means that if we use base-10 logarithms and assume that $\alpha = 10^{-6}$, we get the simpler
\begin{equation}
  k = \frac{6}{|\log_{10}(\epsilon)|} - 1 .
  \label{eq-onechunk-special}
\end{equation}
For example, if the per-chunk error rate is ten percent ($\epsilon = 0.1$), then $|\log_{10}(\epsilon)| = |\log_{10}(1/10)| = 1$, and so $k = 6/1 - 1 = 5$ parities are needed. If instead the per-chunk error rate is just one percent ($\epsilon = 0.01$), then only $k = 6/2 - 1 = 2$ parities are necessary.

In particular, for the same per-chunk error rates as in Table~\ref{tbl:levels}, we get:
\begin{table}[!ht]
\begin{center}
\begin{tabular}{|l|r|r|r|}
\hline
\multicolumn{1}{|c|}{\multirow{2}{1.5cm}{\centering security\\level}} &
\multicolumn{1}{|c|}{\multirow{2}{1.5cm}{\centering error\\rate}} &
\multicolumn{1}{|c|}{\multirow{2}{1.5cm}{\centering parities\\required}} &
\multicolumn{1}{|c|}{\multirow{2}{1.6cm}{\centering dispersed\\replicas}}\\&&&\\\hline\hline
\textsc{none} & 0\%     & 0 & 0  \\
\textsc{medium} &   1\% & 2 & 2\\
\textsc{strong} &   5\% & 4 & 4 \\
\textsc{insane} &   10\% & 5 & 8 \\
\textsc{paranoid} & 50\% & 19 & 16\\
\hline
\end{tabular}
\end{center}
\caption{For a given per-chunk error rate (first column), how many parities (second column) are required of a single chunk to keep the overall data corruption probability below $\alpha = 10^{-6}$?}
\label{tbl:single-chunk}
\end{table}

\section{Prefetching strategies for retrieval}
\label{sec:strategies}

When downloading, systematic per-level erasure codes allow for different \emph{prefetching strategies}:
\begin{labelledlist}
\item[\textsc{NONE} = \emph{direct with no recovery; frugal}] No prefetching takes place, RS parity chunks are ignored if present. Retrieval involves  only the original chunks, no recovery. 
\item[\textsc{DATA} = \emph{prefetching data but no recovery; cheap}] Prefetching data-only chunks, RS parity chunks are ignored if present, no recovery.
\item[\textsc{PROX} = \emph{distance-based selection; cheap}] For all intermediate chunks, first retrieve $ m$ chunks that are expected to be the fastest to download (e.g., the $m$ closest to the node).
\item[\textsc{RACE} = \emph{latency optimised; expensive}] Initiate requests for all chunks within the scope (max $m+k$) and will need to wait only for the first $m$ chunks to be delivered in order to proceed. This is equivalent to saying that the  $k$ slowest chunk retrievals can be ignored, therefore this strategy is optimal for latency at the expense of cost.
\end{labelledlist}

All in all, strategies using recovery  can effectively overcome the occasional unavailability of chunks, be it due to faults such as network contention, connectivity gaps in the Kademlia table, node churn, overpriced neighbourhoods, or even malicious attacks targeting a specific neighbourhood. 

Similarly, given a typical model of network latencies for chunk retrieval, erasure codes in \textsc{RACE} mode can guarantee an upper limit on retrieval latencies.%
\footnote{For instance, in the temporally sensitive case of real-time video streaming, for any quality setting (bitrate and FPS), buffering times can be guaranteed if the batch of chunks representing a time unit of media is encoded using its own scope(s) of erasure coding.}

\section*{Acknowledgements}

We thank Andrea Robert for her comments and thorough editing work which have greatly improved the paper.


\end{document}